\documentclass[reprint, amsmath, amssymb, aps, prb, superscriptaddress, showkeys,floatfix,]{revtex4-2}
\usepackage[utf8]{inputenc}
\usepackage{graphicx}
\usepackage{dcolumn}
\usepackage{bm}
\usepackage{soul}
\usepackage{color}
\usepackage{mathtools}

\usepackage{makecell}

\DeclareUnicodeCharacter{2264}{}
\DeclareUnicodeCharacter{25A1}{}
\DeclareUnicodeCharacter{2A7D}{}
\usepackage{multirow}
\usepackage[colorlinks=true, allcolors=blue]{hyperref}
\begin{document}

\title{Electrical- and magneto-transport across the thermo-elastic martensitic transformation in anti-site-disordered off-stoichiometric Co-Fe-Ti-Si \\Heusler alloy thin films}

\author{Mainur Rahaman}
\affiliation{School of Physics, University of Hyderabad, Hyderabad - 500046, Telangana, India}

\author{Lanuakum A Longchar}
\affiliation{School of Physics, University of Hyderabad,  Hyderabad - 500046, Telangana, India}

\author{Rajeev Joshi}
\affiliation{UGC - DAE Consortium for Scientific Research, Indore-452001, India}

\author{R. Rawat}
\affiliation{UGC - DAE Consortium for Scientific Research, Indore-452001, India}

\author{M. Manivel Raja}
\affiliation{Defence Metallurgical Research Laboratory, Hyderabad-500058, Telangana, India}

\author{S. N. Kaul}
\affiliation{School of Physics, University of Hyderabad, Hyderabad - 500046, Telangana, India}

\author{S. Srinath}
\email{srinath@uohyd.ac.in}
\affiliation{School of Physics, University of Hyderabad, Hyderabad - 500046, Telangana, India}

\begin{abstract}
Anti-site disorder (ASD), present in Heusler alloy thin films, is known to significantly affect their physical properties, but a complete understanding of the actual role of ASD is still lacking. In this work, we systematically investigate the effect of ASD on electrical resistivity $\rho(T)$ and transverse magnetoresistance $\mathrm{MR}_{\perp}$ in off-stoichiometric Co--Fe--Ti--Si (CFTS) thin films across the thermo-elastic martensitic phase transformation (MPT). The CFTS films with A2 ASD exhibit a negative temperature coefficient of resistivity (n--TCR) and an upturn below $\sim 30\,$K. In sharp contrast, the partially L2$_1$-ordered films are metallic in nature, characterized by a resistivity minimum at low temperatures ($T_{\min} \cong 30\,$K) and a positive TCR for $T > T_{\min}$. The change in the sign of TCR finds a straightforward explanation in terms of the competition between the quantum corrections (weak localization, electron--diffuson scattering) and the ballistic scattering mechanisms (electron--magnon, $e - m$, and electron--phonon, $e-p$). We find that, stronger the atomic ASD, more prominent the quantum corrections and the weaker the scattering of $e-m$ and $e-p$ scattering. All the CFTS films exhibit a distinct thermal hysteresis and a significant drop in resistivity, symptomatic of a MPT, near the characteristic temperatures: martensite-end $T_{\mathrm{Me}} \cong 300\,$K and austenite-begin $T_{\mathrm{Ab}} \cong 325\,$K. Regardless of the strength of ASD, in the martensite phase the anti-symmetric (ASMR) component of $\mathrm{MR}_{\perp}(H)$ dominates over the symmetric (SMR) counterpart, whereas the reverse is true (i.e.\ $\mathrm{SMR} \gg \mathrm{ASMR}$) for the austenite phase at temperatures $T_{\mathrm{Ab}} \cong 325\,$K $\le T \le 375\,$K, where $\mathrm{MR}_{\perp}$ increases very sharply with temperature as the austenite phase grows rapidly at the expense of the martensite phase. The present results assert that the CFTS Heusler alloy thin films are promising candidates for shape-memory devices and for spintronic applications such as spin valves.


\end{abstract}

\maketitle

\section{\label{sec:level1}INTRODUCTION}

Heusler alloys, a class of intermetallic compounds, have garnered significant attention due to their multifunctional properties such as half-metallicity (HM), Spin gapless semiconductor behavior (SGS), high spin polarization, large magnetoresistance (MR), tunable magnetic anisotropy, and shape memory effect (SME), which are crucial for next-generation spintronics applications \cite{elphick2021,dieny2020,rani2020,wang2020,parkin2004,yu2015}. Among them, the SME associated with martensitic phase transformations (MPT) has been extensively explored in Ni-based Heusler alloys \cite{ullakko1996, ma2010,banik2008,kaul2006,barandiaran2009,blinov2020, granovskii2012,blinov2020_2,chatterjee2010,samanta2014} for its applications in magnetic sensors and refrigeration technologies. The MPT induces internal strain, leading to the formation of several martensitic variants within the ferromagnetic material. 

While extensive research has been conducted on Ni-based Heusler alloys \cite{ullakko1996, ma2010,banik2008,kaul2006,barandiaran2009,blinov2020, granovskii2012,blinov2020_2,chatterjee2010,samanta2014} for their SME and magnetic field-induced strain effects, recent investigations on Fe- \cite{omori2009,zhu2009} and Co-based Heusler compounds \cite{roy2016,wuttig2001,oikawa2006,liu2023,hirata2015,xu2017,jiang2019,jiang2017,nakamura2021,li2019} have further advanced the understanding of martensitic phase transitions (MPT) in these systems. Notably, MPT has been identified in off-stoichiometric Co-Ni-Ga \cite{wuttig2001,oikawa2006}, Co-Fe-V-Ga \cite{liu2023}, Co$_2$Cr(Ga,Si) \cite{hirata2015}, Co$_2$V(Ga/Al/Si) \cite{xu2017,jiang2019,jiang2017}, and Co–V–(Al, Ga, In) \cite{nakamura2021,li2019} alloys. In these Co based systems, the MPT occurs near the ferromagnetic transition temperature and is confirmed through the thermo-mechanical properties, leaving a huge research gap about their electrical and magneto transport properties across the magnetic and structural transition. MPT, observed in many Ni-based SM alloys \cite{banik2008,kaul2006,barandiaran2009,blinov2020, granovskii2012,blinov2020_2,chatterjee2010,samanta2014}, manifests itself as a structural transformation, marked by a thermal hysteresis, a sudden drop in electrical resistivity and significant changes in magnetoresistance. However, a comprehensive understanding of the role of anti-site structural disorder in governing martensitic transformations in Co- and Ni-based Heusler alloys remains elusive.

The advent of spin-gapless semiconductors (SGSs) within the realm of half-metallic Heusler-type ferromagnets marks a significant development in semiconductor spintronics \cite{wang2020,rani2020}. These innovative materials are characterized by a unique electronic band structure, featuring a semiconducting bandgap for minority spin carriers while allowing for a zero bandgap for majority spins at the Fermi level \cite{rani2020}. The unique band structure of SGSs in Heusler alloys leads to several remarkable properties, including high spin polarization (up to 100\%) of both electrons and holes, high carrier mobility, and the ability to switch between n-type and p-type spin-polarized conduction by applying an external field \cite{ouardi2013,ouardi2019erratum,buckley2019,shahi2022,galanakis2014,fu2019,chen2022,kharel2015,mishra2023,du2013}. Small negative temperature coefficient of resistivity (n-TCR) $(\sim -10^{-8}~\Omega\mathrm{cm/K})$, positive linear magnetoresistance and change in carrier type with temperature are the proposed experimental characteristics for SGS, observed in many inverse and quaternary Heusler compounds \cite{ouardi2013,ouardi2019erratum,buckley2019,shahi2022,galanakis2014,fu2019,chen2022,kharel2015,mishra2023,du2013}.

In this context, the off-stoichiometric Co-Fe-Ti-Si (CFTS) quaternary Heusler alloy presents a promising platform for studying the effect of anti-site disorder on MPT and spin gapless semiconductor behavior due to the presence of large half-metallic band gap at Fermi level with negative anisotropic magnetoresistance (AMR) ratio, large saturation magnetization, $M_S = 4 \mathrm{\mu}_B/f.u.$, and Curie temperature, $T_c$ as high as 800 K \cite{chen2017,jin2017}. Our previous ferromagnetic resonance (FMR) and magneto-optic Kerr effect (MOKE)   \cite{rahaman2022,rahaman2024} investigations on off-stoichiometric Co–Fe–Ti–Si (CFTS) Heusler alloy thin films have demonstrated that decrease in anti-site disorder (ASD) plays a crucial role in improving the key physical parameters, including a low Gilbert damping constant, high saturation magnetization, large spin-wave stiffness constant, and significant uniaxial anisotropy energy density.

In this work, we systematically investigate the influence of anti-site disorder on the magnetic, electrical- and magneto-transport properties of off-stoichiometric CFTS thin films, particularly in the context of martensitic phase transformation and SGS behavior. The manuscript is organized as follows. The experimental details, including the thin-film growth conditions and structural characterization of CFTS, are presented in Section II. Section~III describes the experimental results, data analysis, and discussion. Section~III consists of three subsections, of which Section~III.A deals with the electrical resistivity, while Section~III.B discusses the magnetoresistance. Finally, the important conclusions drawn from this work are summarised in Section~IV.

\section{Experimental Details}

CFTS thin films of 100 nm thickness were grown on Si (100) substrate by co-sputtering the stoichiometric Co$_{2}$FeSi (CFS) and Co$_{2}$TiSi (CTS) alloy targets in an ultrahigh vacuum magnetron sputtering chamber at different substrate temperatures ($T_S$) 200°C, 350°C, 450°C, 500°C and 550°C. Based on the deposition temperature, these films are labeled as TS200, TS350, TS450, TS500, and TS550. The actual composition of the film, Co$_{2.00(4)}$Fe$_{0.52(3)}$Ti$_{0.25(2)}$Si$_{1.23(2)}$, was determined by energy dispersion x-ray spectroscopy. The x-ray diffraction spectra demonstrate that the variation in $T_S$ resulted in CFTS films with different amounts of anti-site (AS) atomic disorder. Increasing sharpness of (220) and (422) peaks indicates improvement in the crystallinity in the CFTS films as TS increases from 200°C to 550°C.  The complete absence of (111) and (200) peaks is symptomatic of the $A2$ disorder in relatively low-temperature (200°C and 350°C) deposited CFTS films (TS200 and TS350), whereas the presence of (111) peak confirms the partial $L2_1$ order in TS450, TS500, and TS550 CFTS films.  The deposition conditions and structural characterization were discussed in detail in our earlier reports \cite{rahaman2022,rahaman2024}. ‘Zero-field’ electrical resistivity, $\rho(H = 0)$, and ‘in-field’ resistivity, $\rho(H = 90\ kOe)$, as functions of temperature, were measured by the four-probe method, while maintaining the cooling and heating rates at $\sim 1\, K/min$. Electrical resistivity, $\rho $, versus magnetic field isotherms were recorded at various temperatures, ranging from 10 K to 375 K, on rectangular thin films using Quantum design PPMS. Field (temperature) variation of the transverse magnetoresistance at fixed temperatures ($H = 90$\,kOe) was computed from the $\rho(H)_T$ ($\rho(T)_{H=90\,\mathrm{kOe}}$) data (where the current flows along the length in the film plane and $H$ is applied perpendicular to the film plane) using the relation
\[
\mathrm{MR}_{\perp} = \frac{\rho(H) - \rho(H = 0)}{\rho(H = 0)}\,.
\]
Magnetisation measurements ($M$--$H$ and $M$--$T$) were carried out using a Dynacool PPMS--VSM system.

\section{Results and discussion}
\subsection{Electrical resistivity}
\begin{figure*}
\centering
\includegraphics[scale= 0.8]{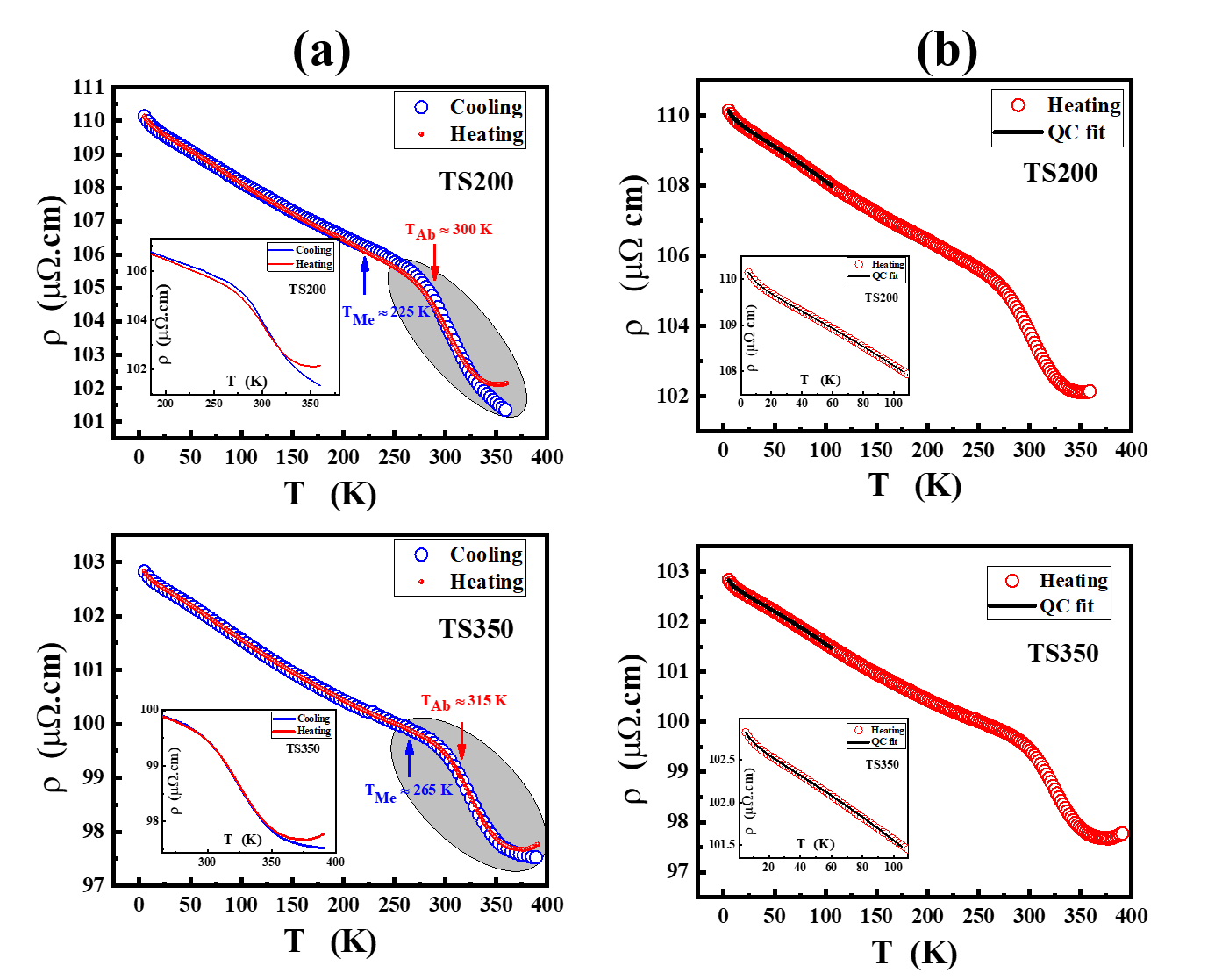}
\caption{(a) Temperature-dependent zero-field resistivity for the disordered TS200 and TS350 CFTS films in the cooling cycle (blue open circles) and the heating cycle (red dots). The insets highlight the regions where thermal hysteresis occurs due to the coexistence of martensite and austenite phases. (b) Zero-field resistivity (red open circles) with theoretical fits (black solid curves) in the temperature range $5$--$105\,$K, based on Eq.~(2) with $\beta_{\mathrm{e\text{-}m}} = \alpha_{\mathrm{e\text{-}p}} = 0$, for the TS200 and TS350 films. The insets provide an enlarged view of the resistivity data together with the theoretical fit in the low-temperature region.
} 
\label{Fig. 1}  
\end{figure*}

\begin{figure*}
\centering
\includegraphics[scale= 0.6]{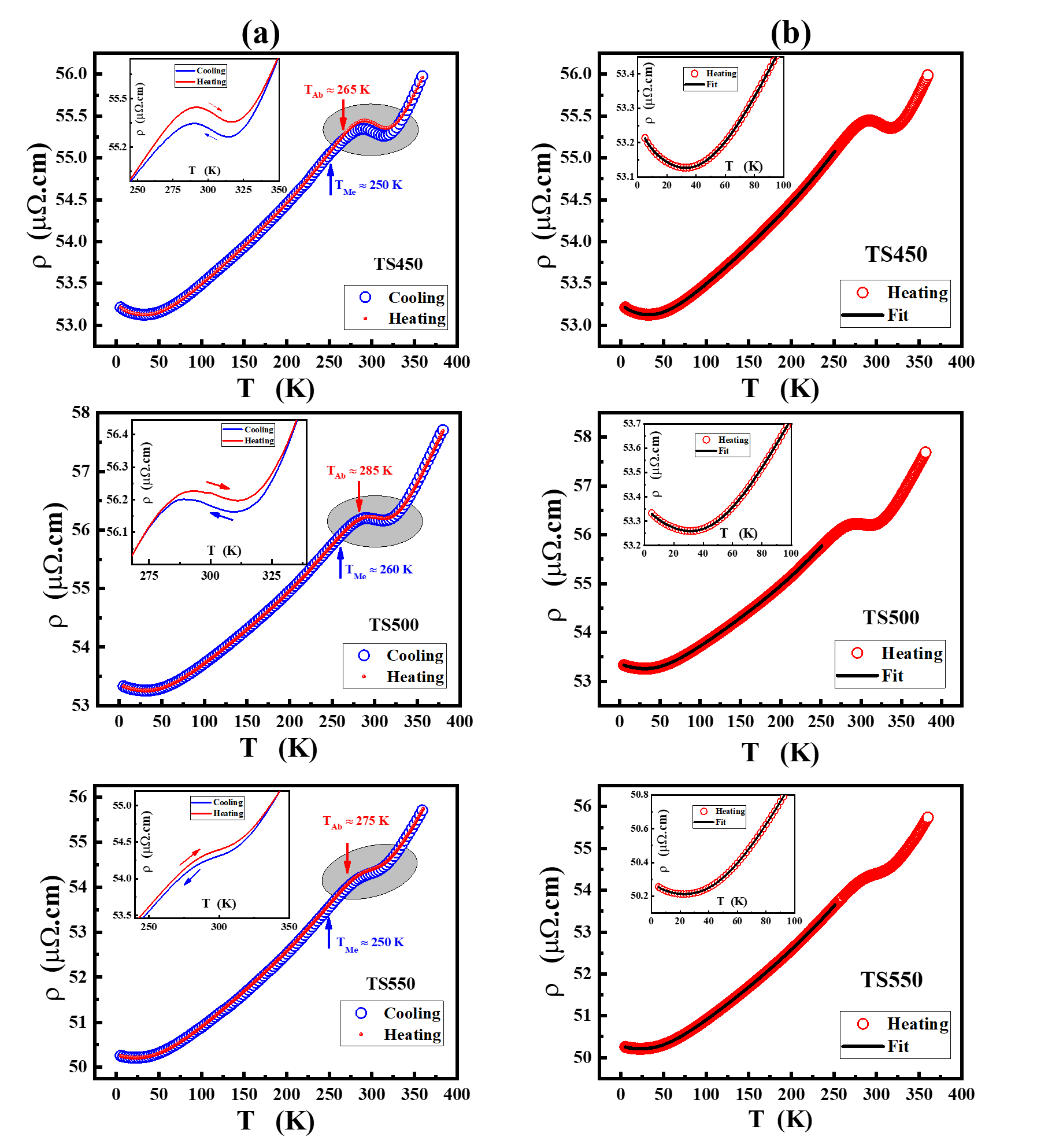}
\caption{(a) Temperature-dependent zero-field resistivity for the ordered TS450, TS500, and TS550 CFTS films in both the cooling cycle (blue open circles) and the heating cycle (red dots). The insets highlight the regions where thermal hysteresis occurs due to the coexistence of martensite and austenite phases.(b) Zero-field resistivity (red open circles) with theoretical fits (black solid curves) in the temperature range $5$--$250\,$K, based on Eq.~(2) for the TS450, TS500, and TS550 films. The insets show an enlarged view of the resistivity data, along with the theoretical fits, in the low-temperature region.
} 
\label{Fig. 2}  
\end{figure*}

Figure \ref{Fig. 1} presents the zero-field resistivity data, $\rho(T, H = 0)$, for the anti-site (A2) disordered TS200 and TS350 CFTS films. An extremely small negative temperature coefficient of resistivity (n-TCR $\sim -10^{-8}\,\Omega\,$cm/K) is observed for $T \le 300\,$K. A n--TCR of such small magnitude has been reported previously in the atomic site-ordered  Mn$_2$CoAl \cite{ouardi2013,ouardi2019erratum,buckley2019,shahi2022,galanakis2014}, CoFeMnSi \cite{fu2019}, CoFeCrAl/Ga \cite{chen2022,kharel2015,mishra2023} and Fe$_2$CoSi \cite{du2013} Heusler alloys and has been identified as a characteristic attribute of spin-gapless semiconductors (SGS). In view of the well-established fact that atomic disorder destroys half-metallicity and SGS behaviour by sustaining a finite density of states in the minority-spin channel at $E_{\mathrm{F}}$, the SGS scenario is not applicable to the present case of the completely site-disordered TS200 and TS350 CFTS films. In sharp contrast, at temperatures above $30\,$K, a typical metallic positive TCR (p--TCR) is observed in the partially L2$_1$-ordered TS450, TS500, and TS550 CFTS thin films, as is evident from figure \ref{Fig. 2}(a).

Considering that a thermal hysteresis between the cooling and heating $\rho(T, H = 0)$ curves (as evidenced in figures \ref{Fig. 1}(a) and \ref{Fig. 2}(a), and highlighted in their insets) is a characteristic experimental signature of the thermo-elastic martensitic phase transformation (MPT), the present results unambiguously assert that all the CFTS films, irrespective of the amount of atomic disorder present, undergo a MPT. Such a resistivity drop and thermal hysteresis near the MPT were observed in many Ni-based Heusler-type shape memory alloys (SMAs) \cite{kaul2006,barandiaran2009,blinov2020, granovskii2012,blinov2020_2,chatterjee2010,samanta2014}. The vertical arrows in figures \ref{Fig. 1}(a) and \ref{Fig. 2}(a) mark the characteristic temperatures for the beginning (end) of the growth of the austenite (martensite) phase at the expense of the martensite (austenite) phase, $T_{\mathrm{Ab}}$ ($T_{\mathrm{Me}}$), while heating (cooling). Thus, all the CFTS films are in the pure martensite phase at temperatures $T \le T_{\mathrm{Me}}$, whereas both martensite and austenite phases coexist at temperatures in the range $T_{\mathrm{Ab}} \le T \le 375\,$K. Note that the Curie temperature $T_{\mathrm{C}}$ ($\sim 800\,$K) of these films lies well above $T_{\mathrm{Me}}$ or $T_{\mathrm{Ab}}$, so that both the martensite and austenite phases remain in the ferromagnetic (FM) state, as is evident from the $M$--$H$ hysteresis loops measured at $T = 5\,$K (martensite phase) and $T = 350\,$K (austenite phase), shown in figure \ref{fig:S1} (see Supplementary material).
To put the above results in a proper perspective, it should be mentioned that, compared to global probes such as magnetization \cite{zuo1998,wang2001,khovailo2004}, strain \cite{cui2004}, and x-ray diffraction \cite{ma2000}, local probes like electrical resistivity are often more reliable for detecting the thermal hysteresis accompanying martensitic transformations, particularly in thin films, because of their sensitivity to local structural changes. Martensitic phase transformations involve changes in the crystal structure that significantly affect electron scattering and hence the resistivity, thereby revealing phase changes at the microscopic level. By contrast, the thermal hysteresis may go undetected in magnetization measurements, especially when the magnetic properties of the ferromagnetic martensite and austenite phases do not differ significantly \cite{barandiaran2009,zuo1998,wang2001,khovailo2004}. In agreement with this premise, figure \ref{fig:S2} of the Supplementary Material highlights the very narrow thermal hysteresis observed in the thermomagnetic data of the CFTS films recorded during heating and cooling cycles at an applied magnetic field of $H = 10\,\mathrm{kOe}$.

Another interesting aspect of the results presented in figures \ref{Fig. 1}(a) and \ref{Fig. 2}(a) is that, in sharp contrast to the ordered CFTS films (TS450, TS500, and TS550), which exhibit a residual resistivity (RR) as low as $50\,\mu\Omega\!\cdot\!\mathrm{cm}$ and a positive temperature coefficient of resistivity (p--TCR), the disordered CFTS films (TS200 and TS350) show a much higher (nearly double) RR of $\cong 110\,\mu\Omega\!\cdot\!\mathrm{cm}$ together with a negative TCR (n--TCR). This suggests that a very short mean free path, arising from substantial disorder, is linked to the negative TCR. The change in the sign of TCR can, in principle, be understood in terms of the Mooij correlation \cite{mooij1973} and its subsequently modified form \cite{tsuei1986}, which empirically suggest that a disordered metallic system with a residual resistivity greater (less) than $\cong 150\,\mu\Omega\!\cdot\!\mathrm{cm}$ generally exhibits n--TCR (p--TCR).

In order to place the temperature-dependent resistivity of both the disordered and partially ordered CFTS films on a consistent theoretical footing, we consider (i) the quantum corrections (QC), such as the enhanced electron--electron interaction ($EEI$), electron--diffuson ($e-d$) scattering, and weak localization ($WL$) effects (which, in disordered metallic systems, dominate at low temperatures and are responsible for the negative TCR), and (ii) the inelastic electron--phonon ($e-p$) and electron--magnon ($e-m$) scattering processes (which dominate at relatively higher temperatures and lead to a positive TCR). A competition between these positive and negative TCR contributions in the partially ordered CFTS films results in a resistivity minimum at a characteristic temperature $T = T_{\min}$.

Assuming the validity of Matthiessen’s rule, the resistivity, $\rho (T)$, in a three-dimensional metallic ferromagnet with weak disorder, can be expressed as

\begin{align}
\rho(T, H) &= \rho_{0} + \rho_{\mathrm{e\text{-}d}}(T) + \rho_{\mathrm{WL}}(T, H) \nonumber \\
&\quad + \rho_{\mathrm{EEI}}(T) + \rho_{\mathrm{e\text{-}m}}(T, H) + \rho_{\mathrm{e\text{-}p}}(T)
\label{eq:1} 
\end{align}

where, $\rho_{0}$ is the residual resistivity. In Eq. \ref{eq:1}, the terms $\rho_{e-d}(T) \sim -\ln(T)$, $\rho_{WL} (T)\sim -T^{(3/2)}$ and $\rho_{EEI} (T)\sim -T^{(1/2)}$ represents, respectively, the electron - diffuson scattering \cite{continentino1978}, weak localization \cite{babu1999,srinivas1999,bergmann1986} and  the enhanced electron-electron interaction \cite{altshuler1985,howson1984,lee1985} contributions to  $\rho (T)$. The remaining two terms in Eq. \ref{eq:1} denote $e-p$ and $e-m$ scattering contribution that give rise to p-TCR \cite{baber1937,ziman1960,wilson1938,raquet2002,kaul2005}. 
The observed $\rho (T)$ data for both disordered and ordered CFTS films are analyzed in terms of Eq. \ref{eq:1}, which includes all possible scattering mechanisms (diffusive as well as ballistic) responsible for electron transport.  Irrespective of the amount of ASD present, the $EEI$ contribution, $\rho_{EEI} (T)$, turns out to be negligibly small. Thus, excluding the $\rho_{EEI} (T)$ term, the explicit form of the expression (Eq. \ref{eq:1}), used in the present case, is given by 

\begin{align}
\rho(T,H) &= \rho_{xx0} - \delta_{dif} \ln T - \xi_{wl} T^{3/2} \nonumber \\
&\quad + \beta_{e-m} \left( \frac{T}{D_{sw}(T)} \right)^2 \nonumber \\
&\quad + \alpha_{e-p} \left( \frac{T}{\theta_D} \right)^3 \int_0^{\theta_D / T} \frac{x^3 e^{x}}{(e^x - 1)^2} \, dx
\label{eq:2}
\end{align}

where $\delta_{dif}$, $\xi_{wl}$, $\beta_{e-m}$, and $\alpha_{e-p}$ are the coefficients of electron-diffuson, weak localization, electron-magnon and electron-phonon scattering contributions, respectively. $D_{sw}(T) = D_0 \left(1 - D_2 T^2\right)$ is the spin wave stiffness, $D_0$ is the value of $D_{sw}$ at 0 K and $D_2$ accounts for the thermal renormalization of $D_{sw}$ due to electron-magnon interactions. $\theta_D$ is the Debye temperature. As the CFTS system consists of 3-d transition metals, magnon-induced interband spin-flip (s↑↓- d↑↓)  scattering \cite{kaul2005,madduri2017,longchar2023a} and phonon-induced non-spin-flip two-band (s↑↓- d↑↓) scattering (known as Block-Wilson model for $e-p$ scattering \cite{wilson1938,longchar2023a,longchar2023b}) are expected to contribute to $\rho(T,H)$, particularly in the partially ordered CFTS films.

Figures \ref{Fig. 1}(b) and \ref{Fig. 2}(b)  clearly demonstrate that the fits, black curves, yielded by Eq. \ref{eq:2} when $\beta_{e-m} = \alpha_{e-p}=0$ (when all the resistivity contributions are present) closely reproduce the $\rho(T)$  data, red open circles, over a large temperature range $5\ K - 105\ K\ (5\ K - 250\ K )$ in the martensite phase,  for the $A2$ disordered (partially $L2_1$- ordered) TS200 and TS350 (TS450, TS500 and TS550) CFTS films. The best fit to the  $\rho(T)$  data are arrived at by following an elaborate procedure detailed in our earlier reports \cite{madduri2017,longchar2023a,longchar2023b}. The optimum values of the fit parameters are tabulated in table \ref{tab:fit_all}.

\begin{table*}
\centering
\caption{Fit parameters for the best fit to the $\rho(T,H=0)$ data based on Eq. \ref{eq:2}}
\label{tab:fit_all}
\vspace{2mm}
\begin{tabular}{|c|c|c|c|c|c|c|c|c|c|}
\hline
CFTS films & 
\makecell{Fit range \\ (K)} &
\makecell{$\rho_{0}$\\($\mu\Omega$cm)} & 
\makecell{$\delta_{\text{dif}}$\\($10^{-2}~\mu\Omega$cm)} & 
\makecell{$\xi_{\text{wl}}$\\($10^{-4}~\mu\Omega$cm~K$^{-3/2}$)} & 
\makecell{$\beta_{e\text{-}m}$\\($10^{-5}~\mu\Omega$cm~K$^{-2}$)} & 
\makecell{$D_2$\\($10^{-6}$~K$^{-2}$)} & 
\makecell{$\alpha_{e\text{-}p}$\\($\mu\Omega$cm)} & 
\makecell{$\theta_D$\\(K)} & 
\makecell{$\chi^2$\\($10^{-6}$)} \\
\hline

TS200 & 5--105 & 110.63(1) & 28.9(2) & 11.4(1) & -- & -- & -- & -- & 2.4 \\

TS350 & 5--105 & 103.07(1) & 14.7(2) & 8.4(1)  & -- & -- & -- & -- & 2.7 \\

TS450 & 5--250 & 53.28(2) & 3.8(1) & 4.0(2) & 1.96(9) & 1.06(8) & 6.98(5) & 363(2) & 1.5 \\

TS500 & 5--250 & 53.37(4) & 2.6(2) & 3.1(1) & 2.4(3) & 1.36(2) & 9.05(8) & 375(2) & 1.9 \\

TS550 & 5--250 & 50.30(2) & 2.7(1) & 2.7(1) & 2.2(1) & 1.37(9) & 9.35(7) & 371(2) & 2.2 \\
\hline

\end{tabular}
\end{table*}

The entries in Table 1 clearly highlight the systematic decrease (increase) in the parameters $\rho_0$, $\delta_{dif}$ and $\xi_{wl}$ ($\beta_{e-m}$, and $\alpha_{e-p}$) in Eq.\ref{eq:2} with diminishing anti-site disorder (ASD). Thus, a direct correlation exists between the atomic ASD and the electron transport mechanisms in the CFTS thin films in question. Disordered films exhibit semiconducting-like behavior caused primarily by the electron-diffuson scattering and weak localization/quantum interference effects. On the other hand, the ordered films prominently display a typical metallic conducting behavior at $T > 30\ K$, which is mainly governed by the electron-magnon and electron-phonon scattering processes. 

\subsection{Transverse Magnetoresistance}

The temperature (field) variation of the transverse magnetoresistance at $H = 90\,\mathrm{kOe}$ (at fixed temperatures in the range $10\,\mathrm{K} \le T \le 375\,\mathrm{K}$ where the martensitic transformation is observed in $\rho(T)$ for the CFTS films TS200, TS350, TS450, and TS500) was computed from the $\rho(T)_{H=90\,\mathrm{kOe}}$ ($\rho(H)_T$) data using the expression

\begin{equation}
\begin{aligned}
\mathrm{MR}_{\perp}(T,H)
= \frac{\rho(T,H) - \rho(T,H=0)}{\rho(T,H=0)} \times 100\% .
\end{aligned}
\label{eq:3}
\end{equation}

\begin{figure*}
\includegraphics[scale= 0.55]{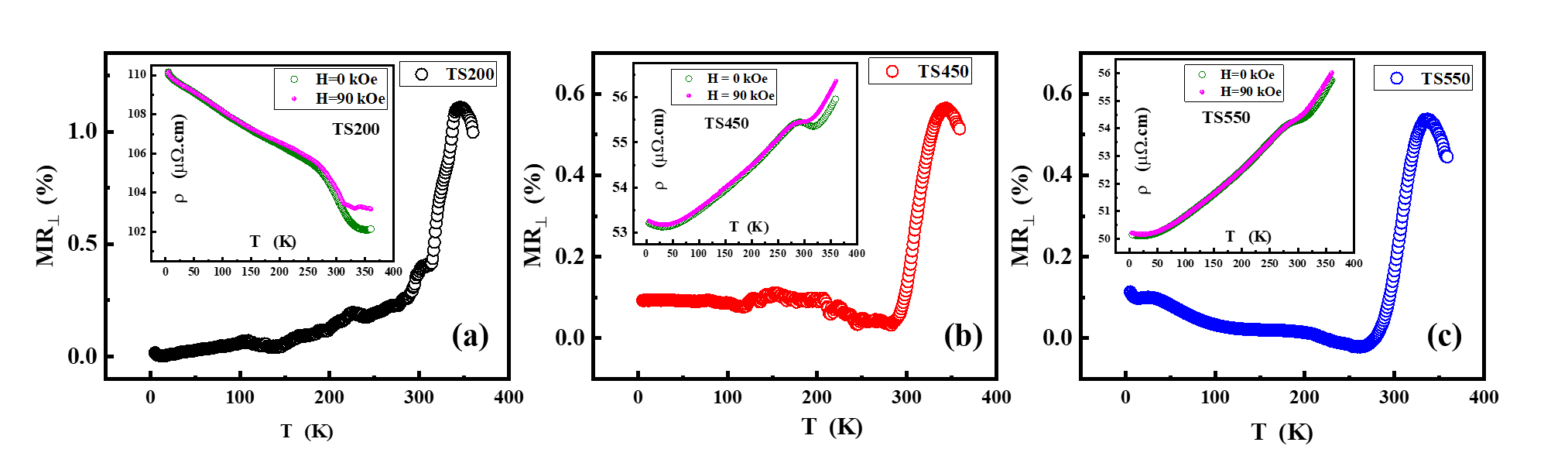}
\caption{$\mathrm{MR}_{\perp}$ as a function of temperature, calculated from $\rho(T)$ at $H = 0$ and $H = 90\,\mathrm{kOe}$ for the CFTS films TS200, TS450, and TS550. The insets display the raw $\rho(T)$ data at $H = 0$ and $H = 90\,\mathrm{kOe}$.
} 
\label{Fig. 3}  
\end{figure*}

The functional dependence of $\mathrm{MR}_{\perp}$ on temperature at $H = 90\,\mathrm{kOe}$ for the CFTS films TS200, TS450, and TS550 is shown in figure \ref{Fig. 3}. This temperature dependence of $\mathrm{MR}_{\perp}$ is representative of the other CFTS films as well. From the data in figure \ref{Fig. 3}, it is evident that in the martensite regime ($T \le T_{\mathrm{Me}} \cong 250\,\mathrm{K}$), $\mathrm{MR}_{\perp}$ remains nearly constant at a very small value of $\sim 0.1\%$, and increases sharply as the temperature rises above $T_{\mathrm{Ab}} \cong 300\,\mathrm{K}$, where the austenite phase grows rapidly at the expense of the martensite phase. The quantity $\mathrm{MR}_{\perp}(T)$ peaks near $T \cong 350\,\mathrm{K}$, attaining values an order of magnitude higher, ranging from $\sim 0.6\%$ to $\sim 1.0\%$, depending on the degree of ASD. Such relatively larger values of $\mathrm{MR}_{\perp}$ are characteristic of the austenite phase.

\begin{figure*}
\centering
\includegraphics[scale=0.63]{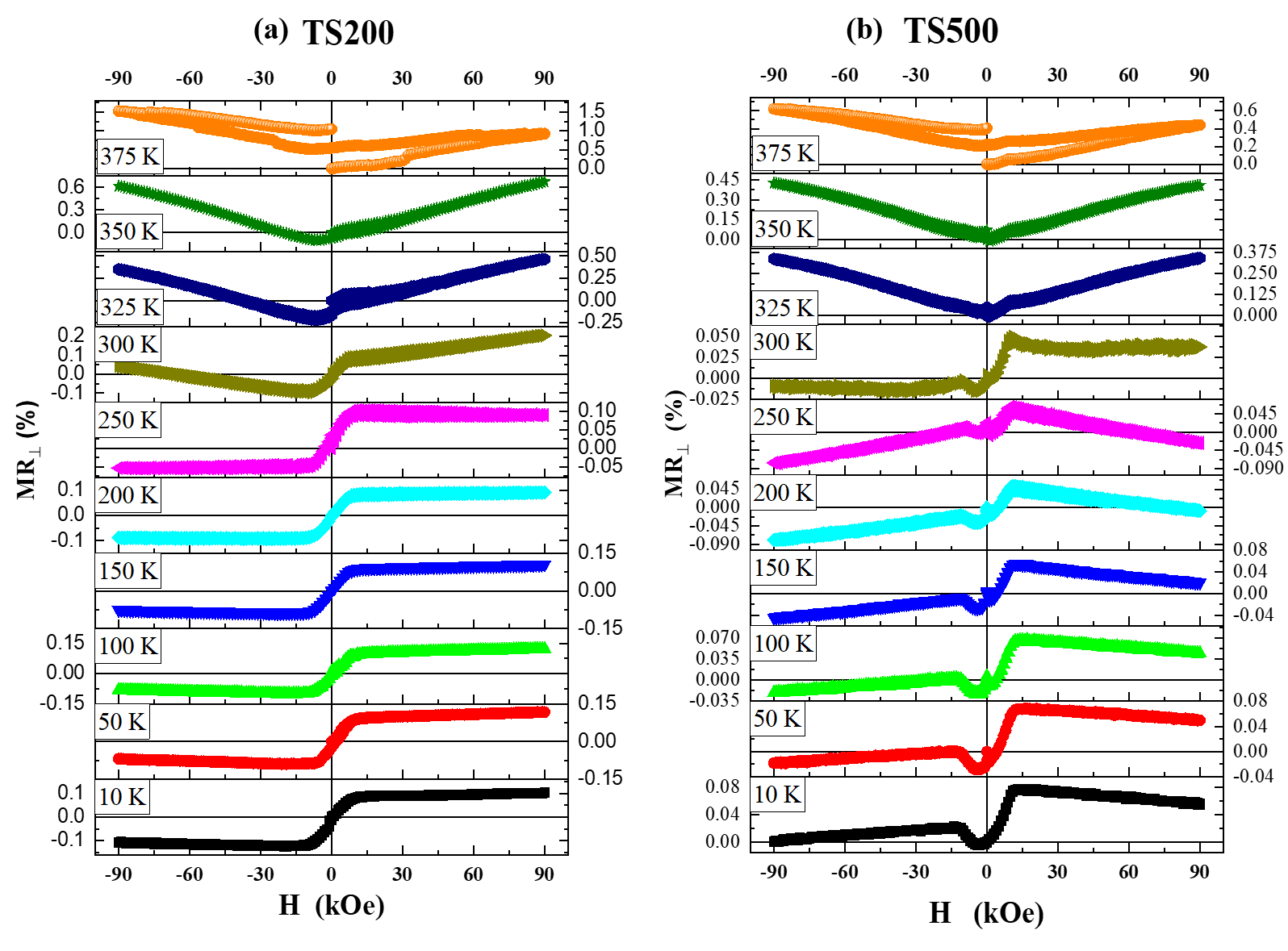}
\caption{$\mathrm{MR}_{\perp}(H)$ isotherms for (a) the completely site-disordered TS200 CFTS film and (b) the optimally site-ordered TS500 CFTS film.
} 
\label{Fig. 4}  
\end{figure*}

Figure \ref{Fig. 4} shows the transverse magnetoresistance $\mathrm{MR}_{\perp}(H)$, calculated from $\rho(H)$ at fixed temperatures for the CFTS films (a) TS200 and (b) TS500. $\mathrm{MR}_{\perp}(H)$ data for TS350 and TS450 are presented in figure \ref{fig:S3} of the Supplementary Material. Irrespective of the strength of the anti-site disorder, for all the films, $\mathrm{MR}_{\perp}(H)$ isotherms demonstrate that, in the martensitic phase region ($10 \lesssim T \lesssim 300~\mathrm{K}$), $\mathrm{MR}_{\perp}$ changes sign with $H$ (i.e., $\mathrm{MR}_{\perp}(H) = -\mathrm{MR}_{\perp}(-H)$ or alternatively, $\mathrm{MR}_{\perp}$ is anti-symmetric with respect to the direction of $H$) and has a very small magnitude ($\mathrm{MR}_{\perp} \sim 0.1\%$). In sharp contrast, at temperatures $T_{\mathrm{Ab}} \cong 325~\mathrm{K} \le T \le 375~\mathrm{K}$, $\mathrm{MR}_{\perp}(H)$ has a relatively higher value (ranging from $\sim 0.6\%$ to $\sim 1.0\%$) at $H = 90~\mathrm{kOe}$ and, except at very low fields ($H \sim 0$), $\mathrm{MR}_{\perp}(H)$ does not change sign with $H$ (i.e., $\mathrm{MR}_{\perp}(H) = \mathrm{MR}_{\perp}(-H)$ or alternatively, $\mathrm{MR}_{\perp}$ is symmetric with respect to the direction of $H$) and increases linearly with $H$. A positive linear magnetoresistance ($\mathrm{MR}_{\perp} \propto H$) of similar magnitude has also been previously reported in L2$_1$ or B2 ordered Mn$_2$CoAl~\cite{kudo2021}, CoFeVSi~\cite{yamada2018} and CoFe(V$_{1-x}$Mn$_x$)Si~\cite{yamada2019} thin films and attributed to a gapless band structure in the minority-spin channel at $E_{\mathrm{F}}$. Now that the spin gapless semiconducting (SGS) mechanism cannot explain the metallic behavior observed in the partially $L2_1$-ordered CFTS thin films, the SGS mechanism is unlikely to be at the root of the positive linear $\mathrm{MR}_{\perp}$ observed in the austenite phase in the present case.

\begin{figure*}
\centering
\includegraphics[scale=0.63]{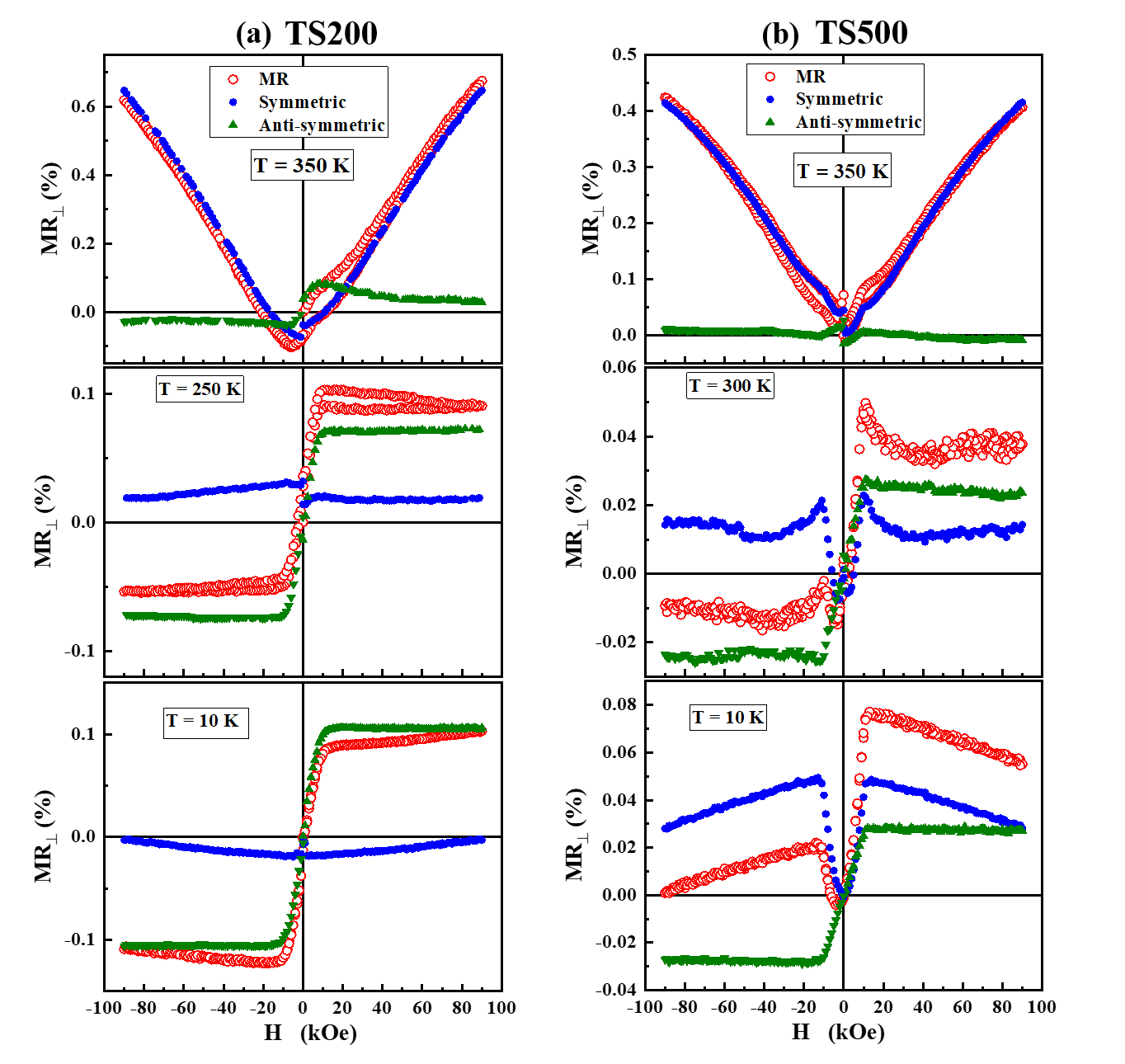}
\caption{Individual symmetric (blue symbols) and anti-symmetric (green symbols) components of the $\mathrm{MR}_{\perp}(H)$ data (red open circles) for (a) the completely site-disordered TS200 CFTS film at $10\,$K, $250\,$K, and $350\,$K, and (b) the optimally site-ordered TS500 CFTS film at $10\,$K, $300\,$K, and $350\,$K.
} 
\label{Fig. 5}  
\end{figure*}

An abrupt change in the slope of linear $\mathrm{MR}_{\perp}$ versus $H$ isotherms in the high-field region ($H \ge 15~\mathrm{kOe}$) is observed at $T_{\mathrm{Me}} \simeq 300~\mathrm{K}$ (Fig.~4). The irreversibility in $\mathrm{MR}_{\perp}(H)$ isotherms at temperatures $T_{\mathrm{Ab}} \simeq 325~\mathrm{K} \le T \le 375~\mathrm{K}$ is indicative of an incomplete martensite-to-austenite phase transformation and the presence of mixed (martensite and austenite) phases at such temperatures.
In order to gain physical insight into the mechanism responsible for the $\mathrm{MR}_{\perp}(H)$ hysteresis loop isotherms, full $\mathrm{MR}_{\perp}(H)$ hysteresis cycle is divided into symmetric and anti-symmetric components using the expressions~\cite{venkateswara2023}

\begin{equation}
\begin{aligned}
    \mathrm{MR}_{\perp}^{\mathrm{Sym.}} &= \frac{\mathrm{MR}_{\perp}(+H^{0 \rightarrow +90 \ \mathrm{kOe}}) + \mathrm{MR}_{\perp}(-H^{0 \rightarrow -90\ \mathrm{kOe}})}{2} \\
\mathrm{MR}_{\perp}^{\mathrm{Asym.}} &= \frac{\mathrm{MR}_{\perp}(+H^{+90\ \mathrm{kOe} \rightarrow 0}) - \mathrm{MR}_{\perp}(-H^{-90\ \mathrm{kOe} \rightarrow 0})}{2}
\end{aligned}
\label{eq:4}
\end{equation}

Figure \ref{Fig. 5} displays the symmetric (blue symbols) and anti-symmetric (green symbols) parts of the $\mathrm{MR}_{\perp}(H)$ data (red open circles) for (a): the completely site-disordered TS200 CFTS film at $10~\mathrm{K}$, $250~\mathrm{K}$ and $350~\mathrm{K}$, and (b): the optimally site-ordered TS500 CFTS film at $10~\mathrm{K}$, $300~\mathrm{K}$ and $350~\mathrm{K}$. Similar curves at $10~\mathrm{K}$, $300~\mathrm{K}$ and $350~\mathrm{K}$ for (a): the site-disordered TS350 CFTS film and (b): the site-ordered TS450 CFTS film, are shown in figure \ref{fig:S4}. From these figures, it is evident that, regardless of the degree of the anti-site disorder present, the anti-symmetric component at all $H$ almost entirely accounts for the observed $\mathrm{MR}_{\perp}(H)$ over the temperature range $10~\mathrm{K} \le T \le 300~\mathrm{K}$ where the martensite phase is prevalent. Note that, in this temperature range, the symmetric part of $\mathrm{MR}_{\perp}$ is negligibly small. Thus, dominant anti-symmetric MR is a characteristic attribute of the martensite phase. The reverse is true (i.e., the symmetric MR dominates over the anti-symmetric counterpart) at temperatures $T_{\mathrm{Ab}} \cong 325~\mathrm{K} \le T \le 375~\mathrm{K}$ where the austenite phase grows at the expense of the martensite phase with increasing temperature and the $\mathrm{MR}_{\perp}(H)$ isotherms mimic the spin-valve type behavior. The only notable exception to this rule is the TS500 CFTS thin film with maximum L2$_1$ order wherein the martensite phase has comparable contributions from both anti-symmetric and symmetric components of MR. As a consequence, the $\mathrm{MR}_{\perp}(H)$ isotherms exhibit spin-valve-like features even in the martensite regime ($10~\mathrm{K} \le T \le 300~\mathrm{K}$). 

\begin{figure*}
\centering
\includegraphics[scale=0.7]{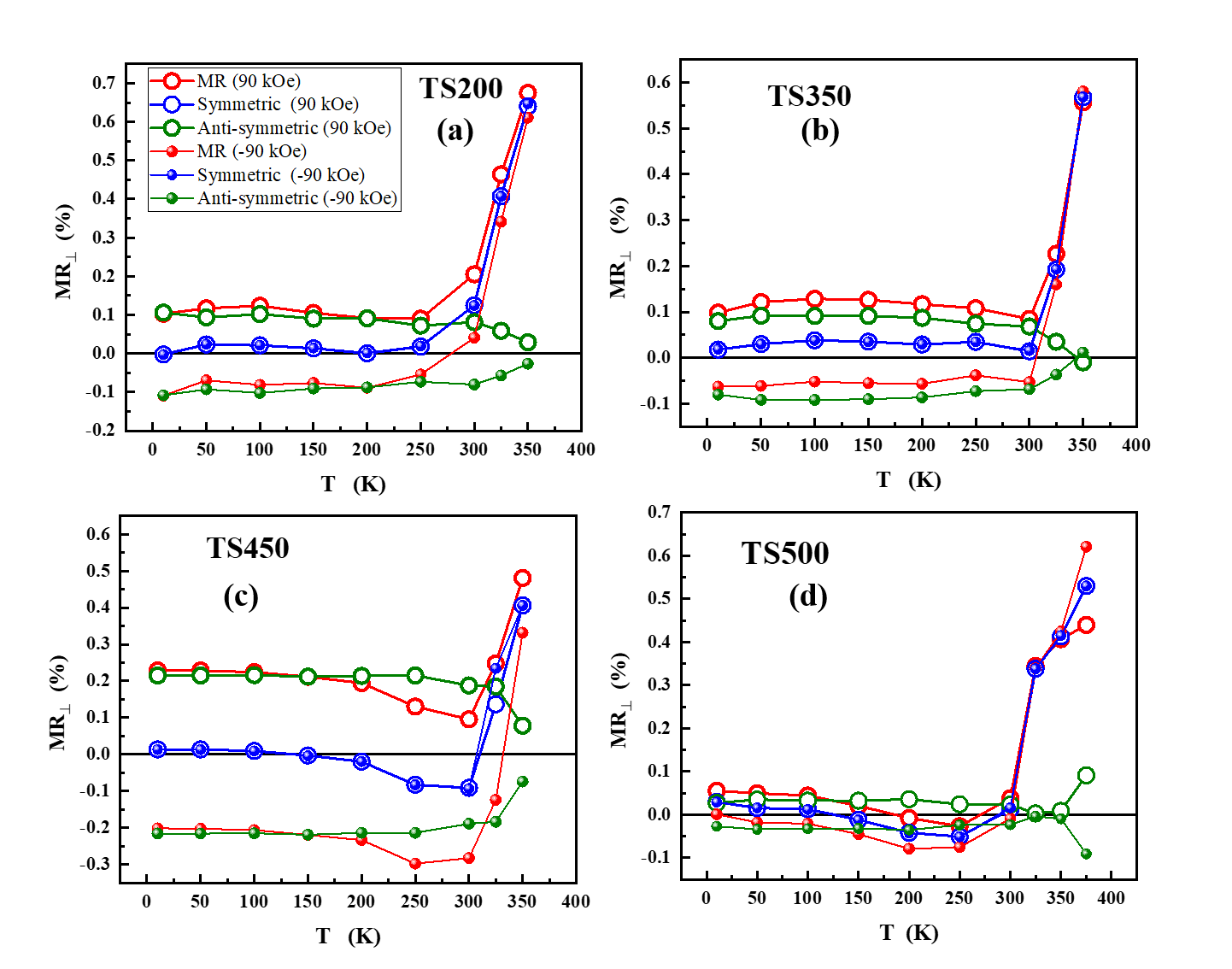}
\caption{Temperature variation of the magnitude of actual observed $\mathrm{MR}_{\perp}(H)$, and its symmetric and anti-symmetric parts extracted at $\pm 90 kOe$ for TS200, TS350, TS450 and TS500 CFTS films} 
\label{Fig. 6}  
\end{figure*}

The temperature variation of $\mathrm{MR}_{\perp}$, observed at $H = \pm 90~\mathrm{kOe}$, and that of the symmetric (SMR) and anti-symmetric (ASMR) parts extracted from the $\mathrm{MR}_{\perp}(H)$ hysteresis loops at $\pm 90~\mathrm{kOe}$, are shown in figure \ref{Fig. 6}. Figure \ref{Fig. 6} clearly demonstrates that, for all the films, $\mathrm{MR}_{\perp}$ is nearly constant at $\sim 0.1\%$ for temperatures up to $T_{\mathrm{Me}} \cong 300~\mathrm{K}$ and increases sharply beyond $T_{\mathrm{Me}}$. Furthermore, in the martensite regime ($10~\mathrm{K} \le T \le T_{\mathrm{Me}} \cong 300~\mathrm{K}$), ASMR dominates over SMR, but the reverse is true (i.e., SMR $\gg$ ASMR) for the austenite phase at temperatures $T_{\mathrm{Ab}} \cong 325~\mathrm{K} \le T \le 375~\mathrm{K}$. These results are in very good agreement with those deduced from the $\mathrm{MR}_{\perp}(T)$ data taken at $H = 90~\mathrm{kOe}$ and shown in Figure \ref{Fig. 3}.

Dominant ASMR (SMR) in the martensite (austenite) phase can be understood as follows. Martensite phase is characterized by a crystal structure of lower symmetry (presumably tetragonal in the present case) compared to the high-symmetry cubic austenite phase. Structural/lattice distortion in the martensite phase gives rise to internal strain and magnetic anisotropy, which break spatial inversion symmetry and/or time-reversal symmetry and enhance spin-dependent electron scattering that is antisymmetric with respect to the magnetic field and hence crucial to ASMR. By contrast, such structural and magnetic effects are negligibly small in the austenite phase and SMR is largely due to the isotropic Lorentz force contribution to MR that involves electronic bands near the Fermi level.

\section{Summary and Conclusions}

The effect of anti-site disorder (ASD) on martensitic phase transformation (MPT) in off-stoichiometric Co--Fe--Ti--Si (CFTS) Heusler alloy thin films is investigated. These thin films were grown on Si(100) substrates at different substrate temperatures ($T_{\mathrm{S}}$) ranging from $200~^{\circ}\mathrm{C}$ to $550~^{\circ}\mathrm{C}$. Increase in $T_{\mathrm{S}}$ results in a change in the degree of anti-site disorder (ASD) from strongly disordered $A2$ (TS200 and TS350 CFTS films) to partially ordered $L2_1$ (TS450, TS500 and TS550 CFTS films). All the films behave as a soft ferromagnet at temperatures ranging from $5~\mathrm{K}$ to $375~\mathrm{K}$.

In all the CFTS film, MPT manifests itself as a distinct thermal hysteresis and a significant drop in resistivity near the characteristic temperatures: martensite-end $T_{\mathrm{Me}} \cong 300~\mathrm{K}$ and austenite-begin $T_{\mathrm{Ab}} \cong 325~\mathrm{K}$. As the magnitude of ASD decreases, the temperature coefficient of resistivity (TCR) in the martensite phase changes from \textit{negative} in the $A2$-disordered CFTS films to \textit{positive} in the partially $L2_1$-ordered CFTS films. The change in the sign of TCR from semiconducting-like to metallic-like, and the occurrence of a resistivity upturn or minimum at low temperatures ($T \cong 30~\mathrm{K}$) have been understood in terms of a competition between the electron-diffuson scattering and weak localization, on the one hand, and the electron-magnon and electron-phonon scattering, on the other. This exercise reveals that, stronger the atomic ASD, more prominent the quantum corrections and weaker the electron-magnon and electron-phonon scattering.

Regardless of the strength of ASD, in the martensite phase, prevalent at temperatures up to $T_{\mathrm{Me}} \cong 300~\mathrm{K}$, the anti-symmetric (ASMR) component of $\mathrm{MR}_{\perp}(H)$ dominates over the symmetric (SMR) counterpart whereas the reverse is true (i.e., SMR $\gg$ ASMR) for the austenite phase at temperatures $T_{\mathrm{Ab}} \cong 325~\mathrm{K} \le T \le 375~\mathrm{K}$ where $\mathrm{MR}_{\perp}$ increases very sharply with temperature as the austenite phase grows rapidly at the expense of the martensite phase.

These findings provide essential insights pertaining to the role of ASD in transport and magneto-transport properties across the MPT, offering promising avenues for designing CFTS-based ferromagnetic shape memory devices. Moreover, the dominance of symmetric $\mathrm{MR}_{\perp}$ in the temperature range $T_{\mathrm{Ab}} \cong 325~\mathrm{K} \le T \le 375~\mathrm{K}$ and ferromagnetic nature of both martensite and austenite phases underscore the potential of these films for spintronic applications such as spin valves.

\section{Acknowledgment}

 MR acknowledges the Department of Science and Technology (DST), India, for the financial support (grant No. DST/ INSPIRE Fellowship/2018/IF180087). SNK is thankful to the Indian National Science Academy (INSA) for funding research under the INSA Honorary Scientist scheme. SS thanks the IOE-UOH for IPDF and also for the sanction (UoH-IoE/SE-IRC/21/01) of Cryogen-free Superconducting Magnetic system for the Magnetic and Transport system.

\bibliography{Reference}
\clearpage
\setcounter{section}{0}
\addcontentsline{toc}{section}{Supplementary Material}

\begin{titlepage}
\centering
    \Large{\textbf{Supplementary Material}}\\[2em]
   \Large{\textbf{Electrical- and magneto-transport across the thermo-elastic martensitic transformation in anti-site-disordered off-stoichiometric Co-Fe-Ti-Si \\Heusler alloy thin films}} \\~\\
    {\large Mainur Rahaman\textsuperscript{1}, Lanuakum A Longchar\textsuperscript{1}, \large Rajeev Joshi\textsuperscript{2}, \large R. Rawat \textsuperscript{2},  \large M. Manivel Raja\textsuperscript{3}, \large S. N. Kaul\textsuperscript{1}, \large S. Srinath\textsuperscript{1, *}}\\[0.4em]
    \textsuperscript{1} \small School of Physics, University of Hyderabad, Hyderabad - 500046, Telangana, India\\
    \textsuperscript{2}UGC - DAE Consortium for Scientific Research, Indore-452001, India \\
    \textsuperscript{3}Defence Metallurgical Research Laboratory, Hyderabad-500058, Telangana, India\\
     \textsuperscript{*}Corresponding author: \texttt{srinath@uohyd.ac.in} \\~\\
\end{titlepage}

\setcounter{figure}{0}
\renewcommand{\thefigure}{S\arabic{figure}}
\setcounter{table}{0}
\renewcommand{\thetable}{S\arabic{table}}

\begin{figure*}
\centering
\includegraphics[scale=0.55]{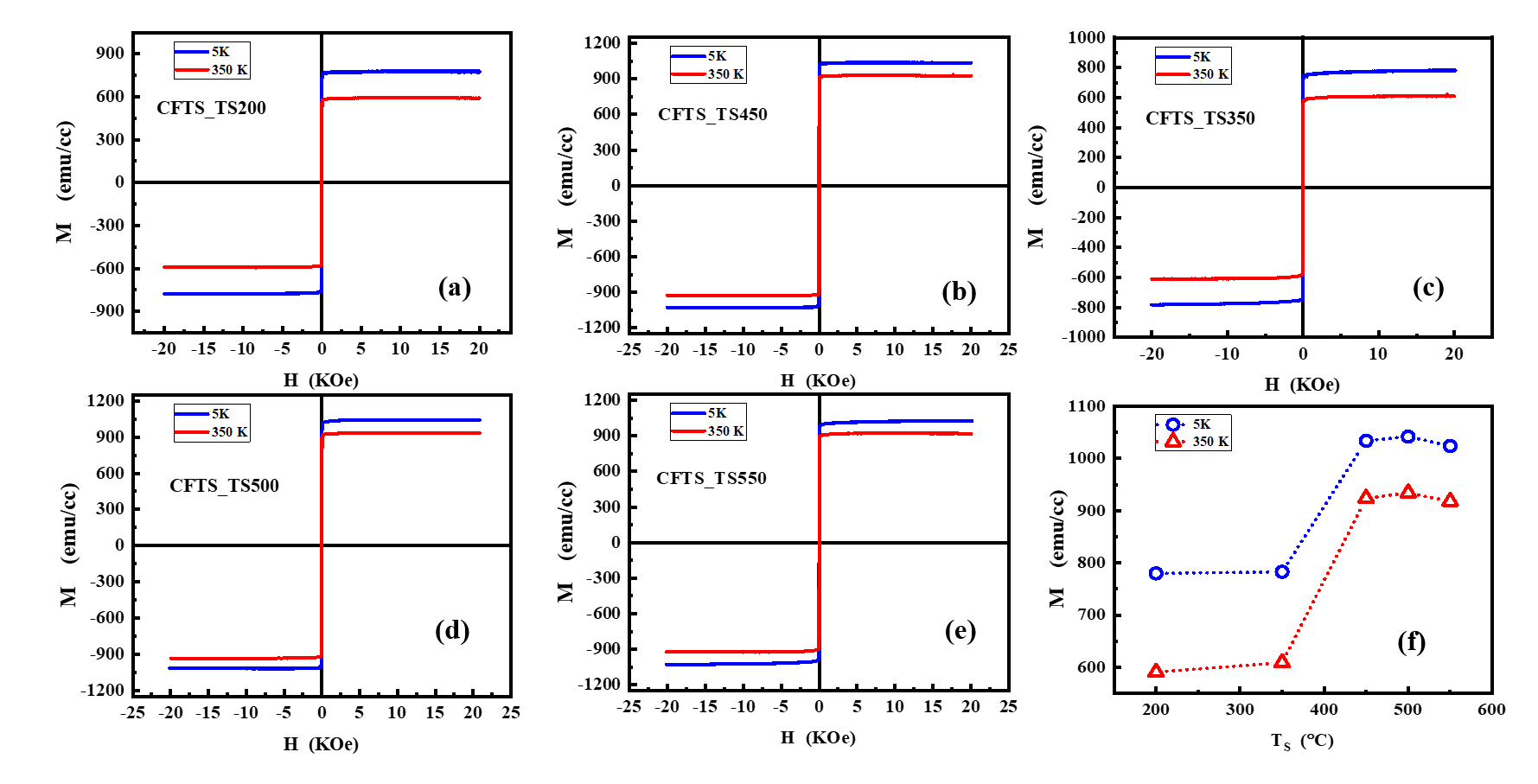}
\caption{(a)--(e) $M-H$ loops measured at $T = 5~\mathrm{K}$ and $350~\mathrm{K}$ for TS200, TS350, TS450, TS500 and TS550 CFTS films. (f): Variation of saturation magnetization with $T_{\mathrm{S}}$.
} 
\label{fig:S1}  
\end{figure*}

$M-H$ hysteresis loops of CFTS films recorded in both the martensite ($T = 5~\mathrm{K}$) and austenite ($T = 350~\mathrm{K}$) phases are shown in Figure \ref{fig:S1}(a)--(e). These results confirm the soft ferromagnetic nature of martensite and austenite phases. The saturation magnetization is higher in the martensite phase compared to the austenite phase for all the films, and it increases with increasing crystalline order and has a maximum value for the TS500 film (Figure \ref{fig:S1}(f)).

\begin{figure*}
\centering
\includegraphics[width=0.9\textwidth]{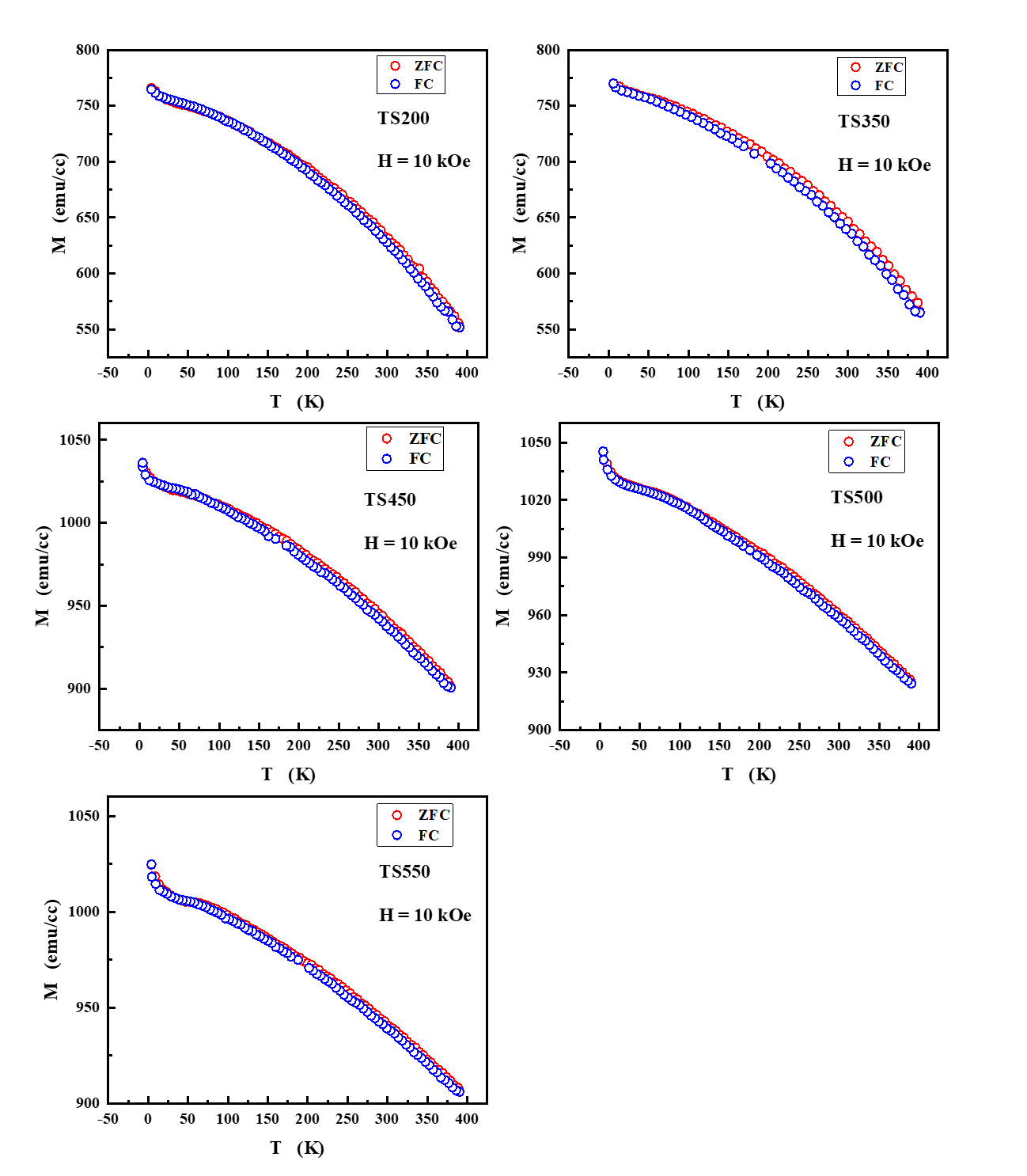}
\caption{$M(T)$ data measured at $H = 10~\mathrm{kOe}$ in both heating (ZFC) and cooling (FC) cycles for TS200, TS350, TS450, TS500, and TS550 CFTS films.
  } 
\label{fig:S2}  
\end{figure*}

Figure \ref{fig:S2} depicts the variation of magnetization with temperature for TS200, TS350, TS450, TS500 and TS550 CFTS films in the heating (`zero-field-cooled`, ZFC, mode) and cooling (`field-cooled`, FC, mode) cycles, recorded at $H = 10~\mathrm{kOe}$. The $M(T)$ data, shown in these subfigures \ref{fig:S2}(a)--(e), exhibit a narrow thermal hysteresis. Contrary to the usual behavior of $M_{\mathrm{FC}}(T) > M_{\mathrm{ZFC}}(T)$, in the present work $M_{\mathrm{ZFC}}(T)$ is consistently higher than $M_{\mathrm{FC}}(T)$ over a wide range of temperatures. A narrow thermal hysteresis is consistent with the fact that both martensite and austenite phases are ferromagnetic in nature and their magnetizations do not differ substantially (figure \ref{fig:S1}).

\phantomsection 

\begin{figure*}
\centering
\includegraphics[width=0.65\textwidth]{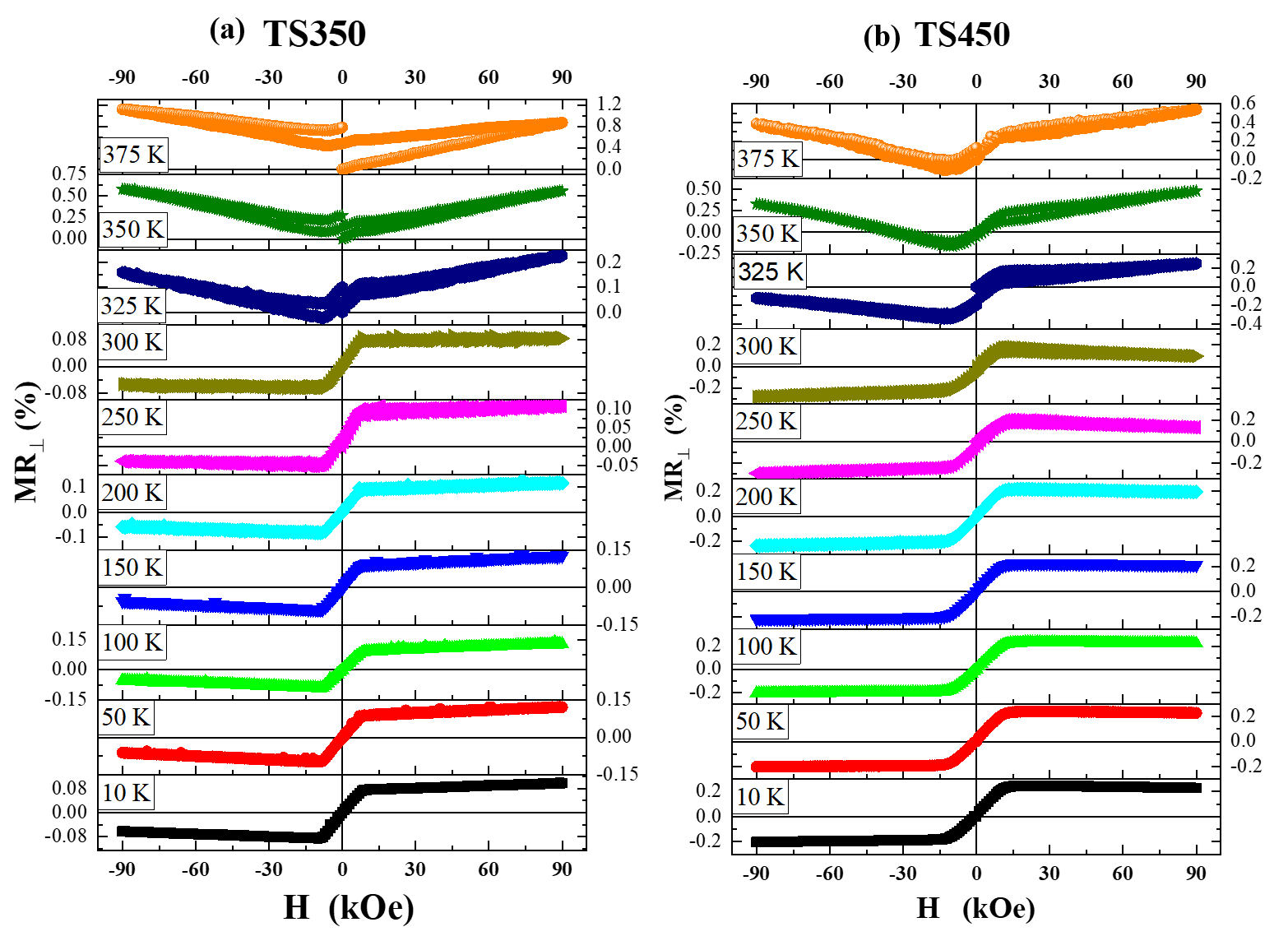}
\caption{(a) $\mathrm{MR}_{\perp}(H)$  calculated at various temperatures for TS350 CFTS film. (b) Individual symmetric (red symbols) and asymmetric (blue symbols) parts of the actual $\mathrm{MR}_{\perp}$  behavior (green symbols) for 10 K, 250K and 350 K.} 
\label{fig:S3}  
\end{figure*}

\begin{figure*}
\centering
\includegraphics[width=0.65\textwidth]{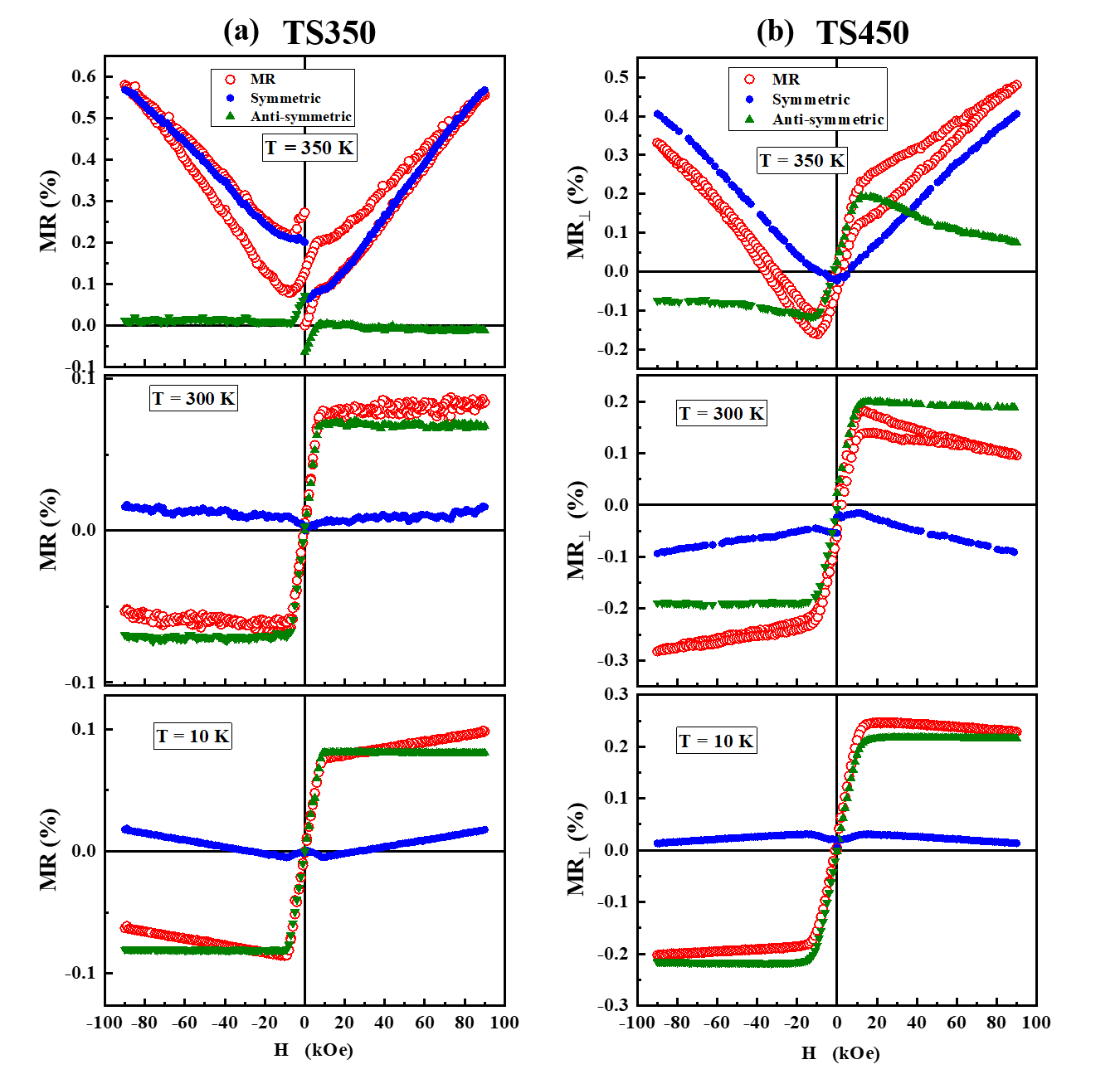}
\caption{(a) $\mathrm{MR}_{\perp}(H)$  calculated at various temperatures for TS450 CFTS film. (b) Individual symmetric (red symbols) and asymmetric (blue symbols) parts of the actual $\mathrm{MR}_{\perp}$  behavior (green symbols) for 10 K, 250K and 350 K.} 
\label{fig:S4}  
\end{figure*}
\end{document}